\documentclass[twocolumn]{jpsj3}

\title{Enhancement of magnetism of Fe by Cr and V}

\author{Masako Ogura, Hisazumi Akai, and Junjiro Kanamori$^{1}$}

\inst{Department of Physics, Graduate School of Science, Osaka University, \\
1-1 Machikaneyama, Toyonaka, Osaka 560-0043, Japan \\
$^{1}$Professor Emeritus, Osaka University, 1-1 Yamadaoka, Suita, Osaka 565-0871, Japan}

\abst{
Enhancement of magnetism of Fe that occurs by alloying with Cr or V is discussed on the basis of first-principles electronic structure calculation.
The $d$ states of Fe next to Cr(V) are pushed down to the lower energy side compared to those of pure Fe due to the hybridization with Cr(V) $d$ states, which leads a Co-like electronic structure for the Fe atom.
This enhances magnetic moments of other Fe atoms and also the exchange couplings among them.
The same mechanism works as well for the magnetism of Fe/Cr heterostructures.
The idea is extended to design of a new type of antiferromagnets with high N{\' e}el temperature.
Such heterostructures could be used to increase a magnetic anisotropy of permanent magnets.
}

\kword{Fe, Cr, V, magnetism, Curie temperature, N{\' e}el temperature, first-principles calculation}

\begin{document}
\maketitle

\section{Introduction}

Most permanent magnets, except for few cases of Co and Mn-based magnets,  are based on iron. Numerous attempts to attain better performance of these iron based permanent magnets have concentrated on enhancing i) the magnetic moment, ii) crystalline  magnetic anisotropy, and iii) Curie temperature. One of common approaches is adding the second and even more ingredients to iron. The most successful one among them is Nd--Fe--B magnet,\cite{sagawa,croat} which is now widely used as a basic component of many products ranging from hard disks to automobiles. Not only due to its world record energy product but also due to its harmo-environmental nature, Nd--Fe--B magnet is recognized as the most important permanent magnet at present. The role of Nd in Nd--Fe--B is to enhance the magnetic anisotropy. This is naively understood as a result of large orbital magnetic moments caused by Nd ions  with the unfilled $f$ shell. On the other hand, the use of B is conceived first  to increase the Curie temperature, though the magnetic anisotropy turns out to be enhanced also by it. To understand the increase of the Curie temperature, it is necessary to examine the effects of the hybridization between Fe $3d$ and B $2p$ states. In short, B transmutes nearby Fe into Co-like atoms. Thus mutated Fe atoms then enhance the magnetism of Fe atoms located further away from those typical element atoms. The same mechanism works not only for B but also for other typical elements such as C and N. This has been long pointed out by one of the present authors (J.K.),\cite{kanamori1, kanamori2, kanamori3} and has also been supported by first-principles calculations\cite{asano} as well as experimental observations.\cite{cadeville} In this paper we assert that this type of mechanism is to work not only for B, C, and N but also for some transition-metal elements, typically V and Cr, and that we may make use of this fact to design a new type of permanent magnets.

The Curie temperature of bcc Fe$_{1-x}$V$_{x}$ ($x \lesssim 0.1$) and Fe$_{1-x}$Cr$_{x}$ ($x \lesssim 0.05$) alloys are higher than that of pure bcc Fe, while their saturation magnetizations are smaller than that of Fe. Although this has been experimentally known for a long time,\cite{bozorth} not much attention has been paid to it so far. Theoretical works using either the first-principles calculations\cite{takahashi} or tight-binding model\cite{kakehashi} on transition metal alloys were performed and they succeeded in reproducing the increase in the Curie temperature. However, to the authors' knowledge, the origin of this increase has not been yet clarified. 

In the present study, we discuss the origin of the enhancement of the magnetism of Fe due to the existence of Cr or V on the basis of first-principles electronic structure calculations. We first explain the enhancement of the magnetism in Fe by Co and by typical elements in \S\ref{sec2}. Our theoretical framework is briefly discussed in \S\ref{sec3}. In the subsequent section, \S\ref{sec4-1}, we demonstrate that a similar mechanism works also for Cr and V.  We propose a design of antiferromagnets with high N{\' e}el temperature that may be useful for developing new permanent magnets in \S\ref{sec4-2}. Finally, in \S\ref{sec5}, we summarizes our study with some concluding remarks.

\section{\label{sec2} A Role of Hybridization in Fe-Based Magnets}

Before entering the discussion of a role of typical elements in ferromagnetism, let us recall the case of Fe$_{1-x}$Co$_x$ alloys that show at $x \sim 0.25$ the highest magnetization per atom among all known ferromagnets.

\begin{figure}
\begin{center}
\includegraphics[scale=0.45]{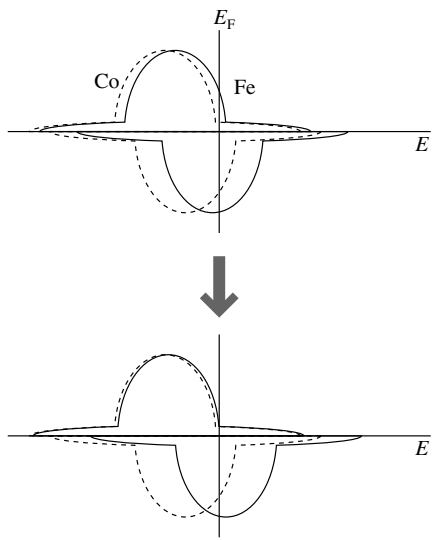}
\end{center}
\caption{Change of the local $d$ state of Fe and Co due to the hybridization.}
\label{sch1}
\end{figure}

The reason why introducing Co into Fe increases its magnetization is the following \cite{HK}. 
The spin-up state (hereafter, the spin-up state is assumed to be the majority spin state) of both Fe and Co is nearly fully occupied. Due to this together with a fact that the intra-atomic Coulomb interaction is comparable to the $d$-band width for $3d$ transition metals, there can be no large potential difference between Fe and Co for spin-up state. On the other hand, for the spin-down state, the atomic potential for Co is deeper than that of Fe, reflecting the difference in the accumulated local $d$-electron numbers as is shown in Fig. \ref{sch1}. This further pushes up and down the local densities of spin-down states of Fe and Co, respectively, through $d$--$d$ hybridization (see Fig. \ref{sch1} (bottom)). This hardly affects the local magnetic moment of Co because the spin-down state is completely filled anyway. However, for Fe, the pushing up of the spin-down state in turn causes a pushing down of the spin-up state, and hence, an increase in the number of spin-up electrons at Fe sites. This is possible because there still remain a small number of $d$-holes in the spin-up state of Fe. For this reason, the local magnetic moment of Fe can increase up to nearly $3\mu_{\rm B}$ in Fe$_{1-x}$Co$_x$ alloys.

\begin{figure}
\begin{center}
\includegraphics[scale=0.45]{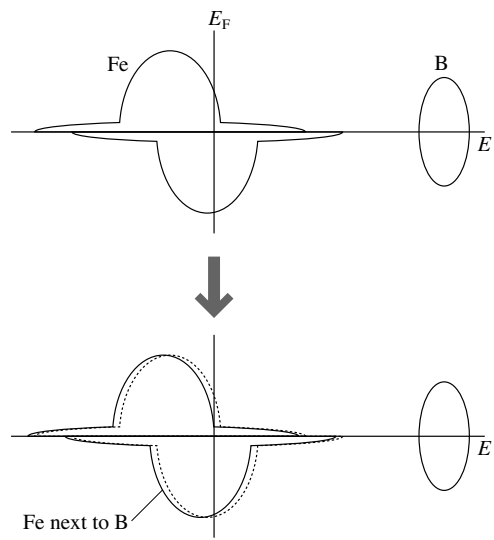}
\end{center}
\caption{Change of the local $d$ state of Fe due to the hybridization with B.}
\label{sch2}
\end{figure}

Now we turn to the cases of typical elements in Fe. The atomic $p$-states of B, C, and N are located at a higher energy region compared to that of Fe $d$-states as is shown in Fig. \ref{sch2} (top) schematically. When these typical elements are introduced into Fe, their $p$-states hybridize with Fe $d$-states. The resulting bonding states that are located at a lower energy region than the energy of the original  Fe $d$-states have its amplitude mainly on the Fe atoms adjacent to the typical elements. Although this does not cause any increases in the local $d$-electron number accumulated at the Fe site, Fe $d$-states thus modified are now quite similar to those of Co (Fig. \ref{sch2} (bottom)).  As was shown through detailed calculation \cite{takahashi}, the exchange coupling between Co and Fe atoms is stronger than that between two Fe atoms. The same is true also for the ``mutated'' Fe and a normal Fe. Thus the magnetization as well as  the Curie temperature can be increased by introducing those typical elements into Fe.

In the following, we will show that a similar situation occurs not only for Fe with typical elements but also for Fe with transition metal elements such as V and Cr.

\section{\label{sec3} Calculation}

The electronic structures were calculated using the Korringa-Kohn-Rostoker Green's function method with the local density approximation of the density functional theory.
We used the muffin-tin potential approximation to describe the electronic potential.
The Fe matrix alloyed with Cr(V) was simulated by use of a supercell method in order to investigate the dependence of the local electronic structure of Fe on the interatomic distance between Fe and Cr(V) atoms.
A bcc structure is assumed as the underlying lattice throughout the calculations.
The lattice constant is fixed to 2.87 \AA.
Neither the lattice expansion nor the lattice relaxation due to the impurities is taken into account.

From the obtained electronic structures, the effective exchange coupling constants $J_{ij}$'s were calculated with the method proposed by Liechtenstein {\it et al}.\cite{liecht}
Thus obtained $J_{ij}$'s were used to construct a Heisenberg Hamiltonian, assuming that a classical spin is located on each ion site.
We obtained the Curie temperature $T_{\rm C}$ by applying the mean field approximation to this model system.

\section{Results and Discussions}
\subsection{\label{sec4-1} Magnetism of Fe with Cr or V}

\begin{figure}
\begin{center}
\includegraphics[scale=1]{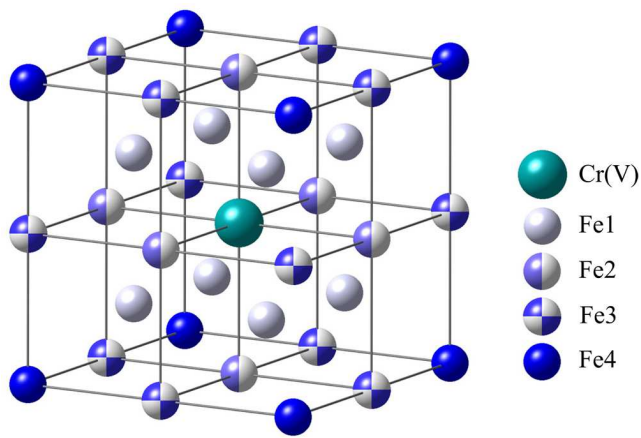}
\end{center}
\caption{(Color online) Supercell of Fe:Cr(V).}
\label{supercell}
\end{figure}

In order to discuss the electronic structure and magnetism of bcc Fe alloyed with Cr or V (Fe:Cr(V), hereafter), we calculated the system shown in Fig. \ref{supercell},
a supercell consisting of fifteen Fe and one Cr(V) atoms.
Here, the nearest Fe atom to the Cr(V) atom is labeled as `Fe1', the second nearest Fe is labeled as `Fe2', and so on.
In this system, both Cr and V atoms couple to Fe antiferromagnetically.

In Table \ref{t1}, the local magnetic spin moments and the saturation magnetization of Fe:Cr and Fe:V are summarized.
For comparison, the magnetic moments of pure Fe are also listed.
Here the local magnetic moment is defined as a magnetic moment within the muffin-tin sphere in each atomic site.
For Fe:Cr, although the local magnetic moment at the Fe1 site is almost the same as that of pure Fe, the magnetic moment becomes bigger as the location is further away from the Cr atom.
This tendency is consistent with observations obtained by neutron scattering experiments.\cite{collins}
The magnetization of Fe:Cr is less than that of pure Fe because Cr has an opposite magnetic moment to that of Fe.

\begin{table}
\caption{Local magnetic spin moment and the saturation magnetization $M_{\rm s}$ of pure Fe, Fe:Cr and Fe:V.}
\label{t1}
\begin{center}
\begin{tabular}{lcc}
\hline
& \multicolumn{1}{c}{Site} & \multicolumn{1}{c}{Moment ($\mu_{\rm B}$/{\rm atom})} \\
\hline
Fe & Fe & 2.33 \\
& $M_{\rm s}$ & 2.30 \\
\hline
Fe:Cr & Fe1 & 2.34 \\
& Fe2 & 2.38 \\
& Fe3 & 2.52 \\
& Fe4 & 2.61 \\
& Cr & -1.80 \\
& $M_{\rm s}$ & 2.10 \\
\hline
Fe:V & Fe1 & 2.26 \\
& Fe2 & 2.29 \\
& Fe3 & 2.43 \\
& Fe4 & 2.56 \\
& V & -1.30 \\
& $M_{\rm s}$ & 2.06 \\
\hline
\end{tabular}
\end{center}
\end{table}

\begin{figure}
\begin{center}
\includegraphics[scale=0.7]{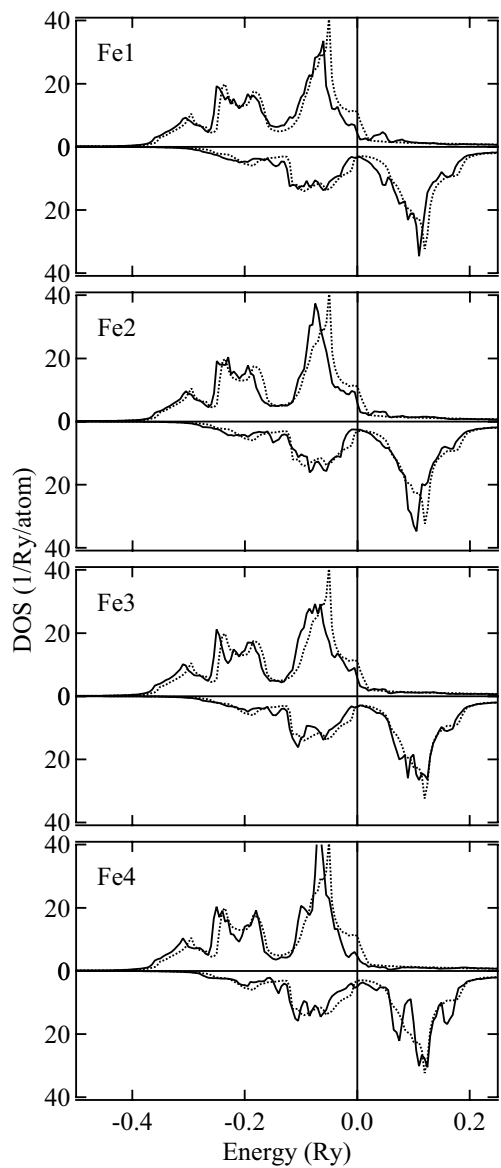}
\end{center}
\caption{The local Fe $d$-DOS of Fe:Cr (solid lines). The spin up and down states are shown in the upper and lower halves, respectively. The dotted lines are the local $d$-DOS of Fe.}
\label{dos}
\end{figure}

In Fig. \ref{dos}, the local $d$ density of states (DOS) at each different Fe site in Fe:Cr is shown.
The local $d$-DOS of pure Fe is also shown in the figure for comparison.
It is seen that, since the Cr $d$ states are located at higher energy region than those of Fe,
the $d$ states at the Fe1 site  are pushed down to the lower energy side compared to those of pure Fe
due to the hybridization between Fe1 and Cr. The situation is quite similar to what is observed
for Fe alloyed with typical elements.
Since both spin-up and down states are pushed down, the local magnetic moment of Fe1 is not affected much.
An important point is that the spin-up state is now almost occupied.
This situation is similar to that of ferromagnetic Co.

Now Fe1 has locally a Co-like electronic structure and  magnetic moments of the other Fe atoms are enhanced by the same mechanism as that explained for Fe$_{1-x}$Co$_x$ alloys in \S\ref{sec2}:
The spin-down states of other Fe sites are pushed up towards the higher energy region as antibonding states because the spin-down $d$ state of Fe1 is located at the lower energy region.
On the other hand, the spin-up state is pushed down because the number of the spin-down electrons decreases.
As a result, the magnetic moments of Fe2, Fe3 and Fe4 increase.
Fe4 has the largest magnetic moment because this site is the furthest from the Cr atom and surrounded by the Fe1 atoms.

\begin{figure}
\begin{center}
\includegraphics[scale=0.65]{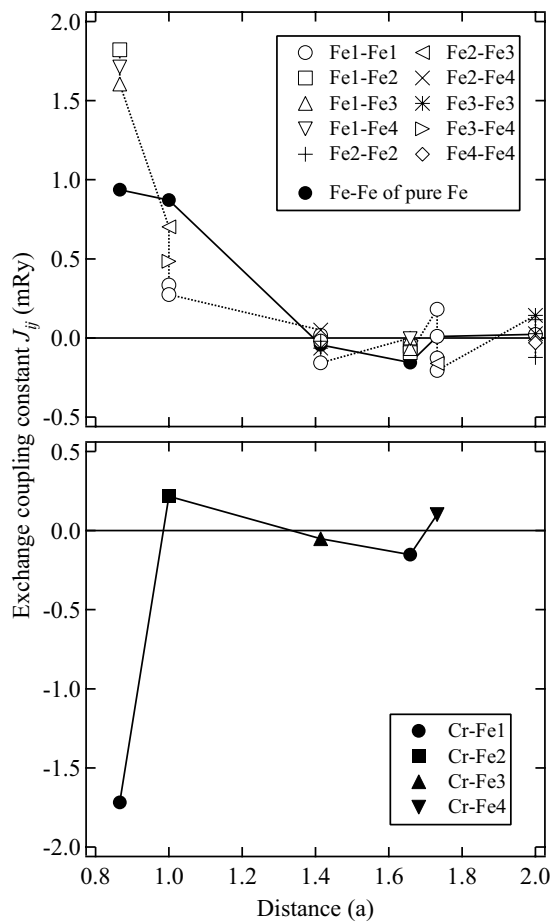}
\end{center}
\caption{Exchange coupling constants of Fe--Fe and Cr--Fe in Fe:Cr. $J_{ij}$'s of Fe--Fe are also shown.}
\label{jij1}
\end{figure}

The change in the electronic structures of Fe due to the existence of Cr atoms also causes the increase in the Curie temperature.
In Fig. \ref{jij1}, exchange coupling constants of pure Fe and Fe:Cr are shown as functions of the atomic distance.
It is clearly seen that $J_{ij}$'s between Fe1 and the other Fe at the distance $(\sqrt{3}/2) a$ are almost doubled from the value of pure Fe.
In addition, $J_{ij}$ between Cr and Fe1, which is at the same distance, is also large.
Although $J_{ij}$'s at the distance $a$ are reduced, the enhancement in $J_{ij}$'s at $(\sqrt{3}/2) a$ prevails the former.
Due to these large exchange couplings, the Curie temperature is enhanced:
$T_{\rm C}$ of Fe:Cr is 1490 K, while that of pure Fe calculated by the same procedure is 910 K.
(Since the experimental $T_{\rm C}$ of pure Fe is 1043 K, the present calculation gives a reasonable estimate for $T_{\rm C}$.)

\begin{figure}
\begin{center}
\includegraphics[scale=0.65]{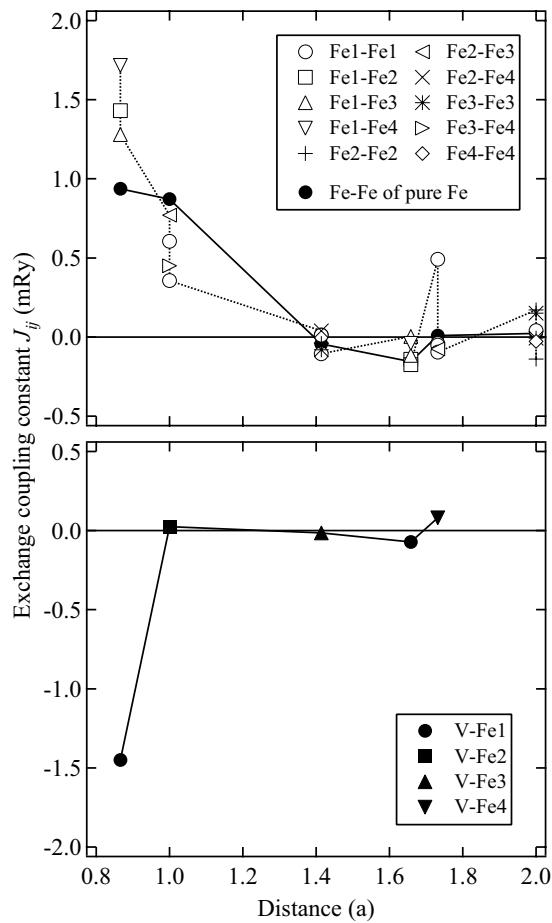}
\end{center}
\caption{Exchange coupling constants of Fe--Fe and V--Fe in Fe:V.}
\label{jij2}
\end{figure}

Fe:V shows a similar behavior to Fe:Cr although the reduction in the local magnetic moment at the Fe1 and Fe2 sites is larger and the enhancement of those at the other Fe sites is smaller than Fe:Cr.
Figure \ref{jij2} shows the exchange coupling constants of Fe:V.
The Curie temperature of Fe:V is also enhanced up to $T_{\rm C}=$ 1270 K.

\subsection{\label{sec4-2} Fe/Cr(V) heterostructures}

It might be difficult to fabricate such compounds as shown in Fig. \ref{supercell}.
However, the increase of the Curie temperature is also attained in the Fe/Cr(V) heterostructures, which can be fabricated experimentally.

\begin{figure}
\begin{center}
\includegraphics[scale=0.7]{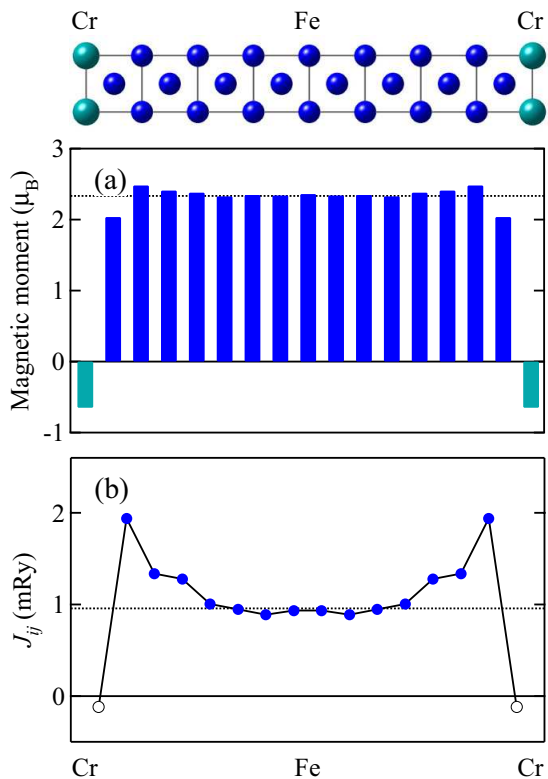}
\end{center}
\caption{(Color online) (a) Distribution of the local magnetic moments in the Fe$_{15}$/Cr heterostructure.
The local moment of pure Fe is indicated by the dashed line.
(b) Distribution of the exchange coupling constants of the nearest pairs in the Fe$_{15}$/Cr heterostructure.  The closed and open circles are $J_{ij}$'s between Fe-Fe and Fe-Cr, respectively. The dashed line is the corresponding $J_{ij}$ in pure Fe.}
\label{layer2}
\end{figure}

In Fig. \ref{layer2}(a), the atomic site-projected local magnetic moments of the Fe$_{15}$/Cr heterostructure are shown.
The magnetic moment of Fe in the nearest layer to the Cr layer is reduced and that in the second nearest layer is enhanced.
Then the magnetic moments decay as the distance from the Cr layer increases, finally converging to the value of pure Fe.

The layer dependent exchange coupling constants between the nearest atom pairs in the Fe$_{15}$/Cr heterostructure are shown in Fig. \ref{layer2}(b).
$J_{ij}$'s between the Fe-Fe pair in the first- and second-nearest layers to the Cr layer are also enhanced as much as the Fe:Cr case.
$J_{ij}$'s also decay as increasing distance from the Cr layer, and converge to the value of pure Fe.
Obtained $T_{\rm C}$ of Fe$_{15}$/Cr heterostructure is 1270 K.
This value is lower than $T_{\rm C}$ of Fe:Cr but considerably higher than that of pure Fe.

\begin{figure}
\begin{center}
\includegraphics[scale=0.7]{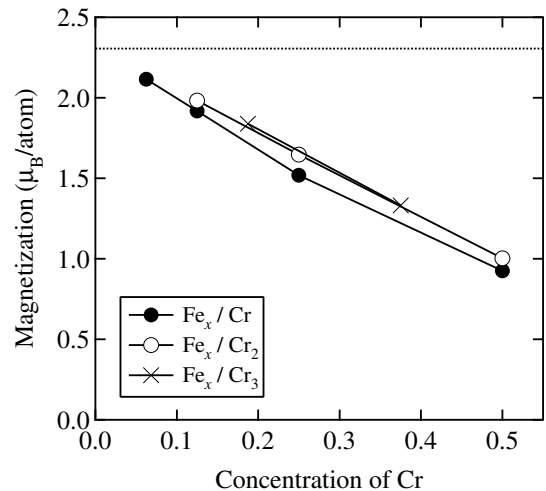}
\end{center}
\caption{Saturation magnetization of the of Fe$_{x}$/Cr$_{y}$ heterostructures. The dashed line is the total magnetic moment of Fe calculated in the same condition.}
\label{moment}
\end{figure}

\begin{figure}
\begin{center}
\includegraphics[scale=0.7]{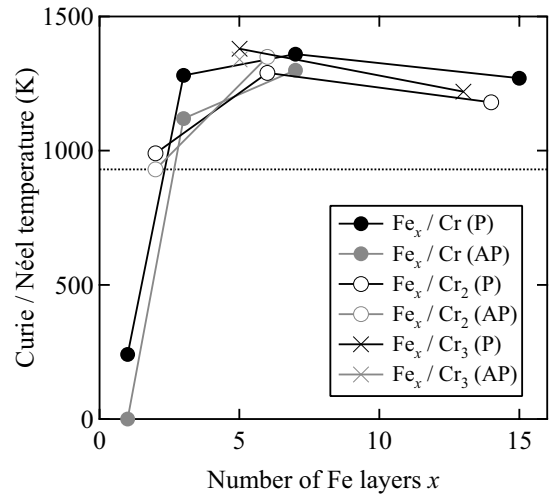}
\end{center}
\caption{Curie/N{\' e}el temperatures of the of Fe$_{x}$/Cr$_{y}$ heterostructures. The results for both parallel (P) and antiparallel (AP) interlayer couplings are shown. The dashed line is the $T_{\rm C}$ of Fe calculated in the same condition.}
\label{tc}
\end{figure}

\begin{figure}
\begin{center}
\includegraphics[scale=0.7]{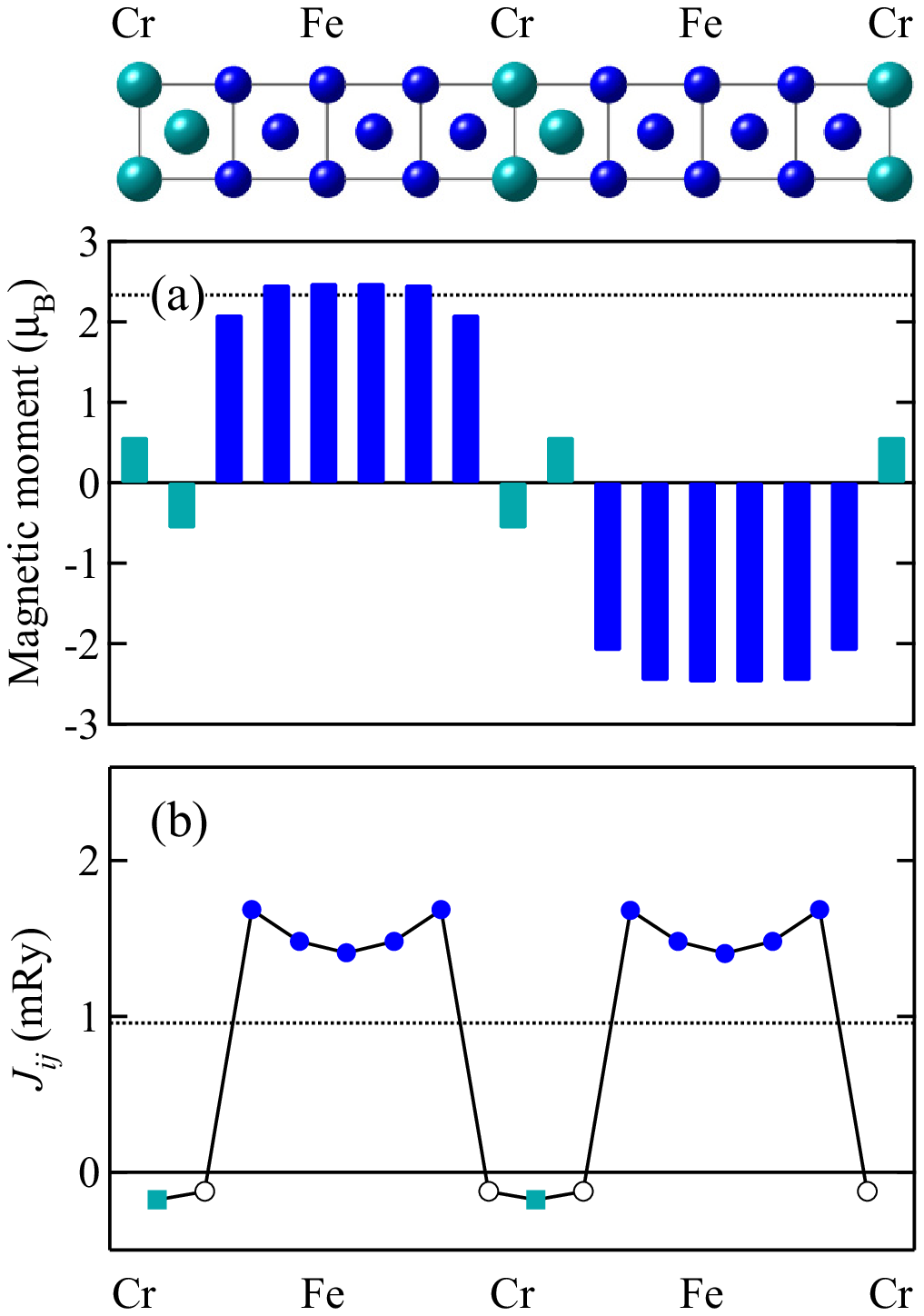}
\end{center}
\caption{(Color online) (a) Distribution of the local magnetic moments in the Fe$_{6}$/Cr$_{2}$ heterostructure (antiparallel interlayer coupling).(b) Distribution of the exchange coupling constants of the nearest pairs in the Fe$_{6}$/Cr$_{2}$ heterostructure.  The squares are $J_{ij}$'s between Cr-Cr.}
\label{layer3}
\end{figure}

The magnetization and magnetic transition temperature of the Fe$_{x}$/Cr$_{y}$ heterostructures are summarized in Figs. \ref{moment} and \ref{tc}, respectively.
Both parallel and antiparallel interlayer magnetic couplings between Fe layers are considered in the present calculation.
When the number of Cr layers $y$ is even, the antiparallel interlayer coupling is more stable than the parallel coupling, and reversed when the number of Cr layers $y$ is odd, due to the antiferromagnetism of Cr.\cite{hirai}
As an example of the antiparallel case, the result for Fe$_{6}$/Cr$_{2}$ is shown in Fig. \ref{layer3}.
Thus the system with even Cr layers is an antiferromagnet whose transition temperature should be called the N{\' e}el temperature.

In Fig.\ref{moment}, the magnetization is plotted as a function of the Cr concentration.
It decreases linearly as increasing Cr concentration.
In Fig.\ref{tc}, the Curie or N{\' e}el temperatures are plotted as a function of the number of Fe layers $x$ in the Fe$_{x}$/Cr$_{y}$ heterostructure.
In contrast to the magnetization, $T_{\rm C}$ of Fe$_{x}$/Cr$_{y}$ takes a maximum at around $x=5$, depending neither on the number of Cr layers $y$ nor the parallel/antiparallel orientation.
As was seen in the previous subsection, Cr does not enhance the magnetism of surrounding Fe but the Fe next to the Cr site enhances the magnetism of the surrounding other Fe atoms.
When the number of Fe layers is small, all Fe atoms are located near Cr and $T_{\rm C}$ cannot increase.
On the other hand, when the number of Fe layers is too large, the number of the Fe sites that are not affected by the Fe next to the Cr atom increases and the increase in $T_{\rm C}$ is prevented.

The antiferromagnets with even Cr layers whose N{\' e}el temperature can exceed the Curie temperature of Fe may offer a new possibility for developing permanent magnets that can be used at high temperatures and free from rare earth elements. The exchange anisotropy mechanism, which was first proposed by Meiklejohn and Bean in 1956\cite{meikle} with a verification by Co particles coated by antiferromagnetic CoO, may provide with a sufficient strength of coercive field for Fe:
The present study predicts that the coupling of the antiferromagnets with ferromagnetic Fe systems may provide with an exchange anisotropy of workable strength.
$J_{ij}$ between Cr and the nearest Fe in the Fe$_{6}$/Cr$_{2}$ heterostructure is -0.178 mRy; this value indicates that Fe atoms on the layer adjacent to Cr layers are subject to an effective field of about 80 T from antiparallel Cr atoms.

The Fe/V heterostructures show a similar behavior to Fe/Cr.
In this case, however, the parallel interlayer coupling is more stable than the antiparallel coupling irrespective of the number of V layers.
This is in accordance with a simple treatment of the exchange interaction between two transition element atoms having magnetic moments based on the Anderson model,\cite{alexander, moriya} which was justified by Hirai and Kanamori for a general assembly of transition elements producing $d$ band states.\cite{hirai2}
The exchange coupling between two neighboring transition element atoms in a metallic system is antiferromagnetic if the total number of $d$ electrons of the two atoms is around 10\cite{kanamori4}, ferromagnetic if the number increases towards 20; the coupling becomes also ferromagnetic towards 0.
Thus the antiferromagnetic V-V coupling in Fe/V is considerably reduced. For this reason the antiferromagnetism is limited to Fe/Cr.

\section{\label{sec5} Summary}

The origin of the enhancement of magnetism in Fe on alloying with Cr or V is investigated on the basis of first-principles electronic structure calculation.
We demonstrated that the mechanism that enhances the magnetism of Fe when it is alloyed with typical elements such as B, C and N also worked for Fe with Cr or V.
Cr and V transmute the neighboring Fe atoms so that they have a Co-like electronic structure.
Such ``cobaltized'' Fe atoms enhance the magnetic moments of further distant Fe atoms as well as the ferromagnetic coupling among them.
This fact explains the increase of the Curie temperature of FeCr and FeV alloys.
The idea was extended to design of a new type of permanent magnets.
It makes use of the exchange anisotropy produced by an antiferromagnetic Fe/Cr heterostrucrure whose N{\' e}el temperature is enhanced by the above mechanism.
Since the epitaxial growth of ferromagnetic Fe layers on the present antiferromagnetic layers along the [001] direction seems to be quite realizable, we hope that our proposal will be examined experimentally.

The concept of the effective filling of the $d$ band or promotion of the atomic number of neighboring transition element atoms by typical elements having 5 or less $sp$ valence electrons can be extended to an understanding of general trends of magnetism in transition metal alloys and compounds. Ferromagnetism in V, Cr, Mn compounds such as V(TCNE)$_{2}\cdot1/2$CH$_{2}$Cl$_{2}$, Cr$_{x}$Mn$_{1-x}$B, CrTe, MnAl, MnB, MnB$_{2}$, MnP, MnAs, etc. is understood at least partially as resulting from the promotion of  V, Cr, and Mn towards Fe, though the $sp$ polarization mechanism of the valence band of typical elements\cite{kanamori5} will be another important driving force in some cases. Another evidence is given by an addition of Al to Cr which enhances the N{\' e}el temperature to indicate a shift of the effective atomic number of Cr towards Mn\cite{akai}. 

\begin{acknowledgments}
We acknowledge Masato Sagawa for fruitful and encouraging discussions.
\end{acknowledgments}

\end{document}